\documentclass[12pt,showpacs,showkeys,preprintnumbers,nofootinbib]{revtex4-1}

\usepackage{amssymb,amsmath,graphics,graphicx}

\begin{document}

\newcommand{\pderiv}[2]{\frac{\partial #1}{\partial #2}}
\newcommand{\deriv}[2]{\frac{d #1}{d #2}}

\title{Inflexibility and independence: Phase transitions in the majority-rule model}

\author{Nuno Crokidakis $^{1}$}
\thanks{E-mail: nuno@if.uff.br}

\author{Paulo Murilo Castro de Oliveira $^{1,2,3}$}
\thanks{E-mail: pmco@if.uff.br}

\affiliation{
$^{1}$Instituto de F\'{\i}sica, Universidade Federal Fluminense, Niter\'oi/RJ, Brazil \\ 
$^{2}$ Universidade Federal da Integra\c{c}\~ao Latino-Americana, Foz do Igua\c{c}\'u/PR, Brazil \\ 
$^{3}$National Institute of Science and Technology for Complex Systems, Rio de Janeiro/RJ, Brazil}

\date{\today}

\begin{abstract}
\noindent
In this work we study opinion formation in a population participating of a public debate with two distinct choices. We considered three distinct mechanisms of social interactions and individuals' behavior: conformity, nonconformity and inflexibility. The conformity is ruled by the majority-rule dynamics, whereas the nonconformity is introduced in the population as an independent behavior, implying the failure to attempted group influence. Finally, the inflexible agents are introduced in the population with a given density. These individuals present a singular behavior, in a way that their stubbornness makes them reluctant to change their opinions. We consider these effects separately and all together, with the aim to analyze the critical behavior of the system. We performed numerical simulations in some lattice structures and for distinct population sizes, and our results suggest that the different formulations of the model undergo order-disorder phase transitions in the same universality class of the Ising model. Some of our results are complemented by analytical calculations.

\end{abstract}



\maketitle

\section{Introduction}

Models of opinion formation have been studied by physicists since the 80's and are now part of the new branch of physics called sociophysics. This recent research area uses tools and concepts of  statistical physics to describe some aspects of social and political behavior \cite{galam_book,sen_book,loreto_rmp}. From the theoretical point of view, opinion models are interesting to physicists because they present order-disorder transitions, scaling and universality, among other typical features of physical systems, which called the attention of many groups throughout the world \cite{lalama,galam_rev,schneider,lccc,martins,martins2,javarone1,javarone2,xiong,sibona,xiong1,xiong2}.

The basic ingredient of models of opinion dynamics is conformity, an important behavior of individuals that emerges as a result of their interactions with other individuals in the population \cite{galam_book,sen_book}. As examples: (i) an individual may copy the state (opinion) of one of his/her neighbors (the voter model \cite{red_book,dornic,roy,mjo_jffm}), or (ii) he/she can consider the majority or the minority opinion inside a small group (the majority-rule models \cite{galam_1999,galam_cont,krap_redner,mjo}), or (iii) a given pair of individuals interact throught kinetic exchanges like an ideal gas \cite{lccc,biswas,biswas2}. Among these models, we highlight the Galam's majority-rule model \cite{galam_1999,galam_cont}. Indeed, influence of majority opinions against minorities have been studied by social scientists since the 50's \cite{asch,jahoda}.

However, recently the impact of nonconformity in opinion dynamics has attracted attention of physicists \cite{sznajd_indep1,sznajd_indep2,sznajd_indep3,nuno_indep}. There are two kinds of nonconformity, namely anticonformity and independence \cite{willis,nail}, and it is important to distinguish between them. The anticonformists are similar to conformists, since both take cognizance of the group norm. Thus, conformists agree with the norm, anticonformists disagree. As discussed in \cite{willis,nail}, an anticonformist actively rebels against influence. This is the case, for example, of the Galam's contrarians \cite{galam_cont}, individuals that known the opinion of the individuals in a group of discussion, and adopt the choice opposite to the prevailing choice of the others, whatever this choice is. On the other hand, we have the independent behavior. In this case, the agent also take cognizance of the group norm, but he/she decides to take one of the possible opinions independently of the majority or the minority opinion in the group \cite{asch,jahoda}. As stated by Willis in \cite{willis}, \textit{``The completely independent person may happen to behave in ways which are prescribed or proscribed by the norms of his group, but this is incidental. It should also be noted that pure anticonformity behavior, like pure conformity behavior, is pure dependent behavior''}.

In terms of the Statistical Physics of opinion dynamics, independence  acts on an opinion model as a kind of stochastic driving that can lead the model to undergo a phase transition \cite{sznajd_indep2,sznajd_indep3}. In fact, independence plays the role of a random noise similar to social temperature \cite{lalama,sznajd_indep2,sznajd_indep3,nuno_indep}. Finally, another interesting and realistic kind of social behavior is usually called inflexibility. Individuals with such characteristic are averse to change their opinions, and the presence of those agents in the population affects considerably the opinion dynamics \cite{martins,moscovici,galam_inflex,jiang,mobilia,galam2011,nuno_celia_victor}. From the theoretical point of view, the introduction of inflexible agents works in the model as the introduction of a quenched disorder, due to the frozen character of the opinions of such agents.

In this work we study the effects of conformity and nonconformity in opinion dynamics. For this purpose, we consider groups of 3 or 5 agents that can interact through the majority rule, but with the inclusion of disorder (inflexibility) and/or noise (independence). We analyze these effects separately in the standard majority-rule model, and all together, in order to study the critical behavior of the system induced by the mentioned effects.

This work is organized as follows. In Section II we present separately in three subsections the microscopic rules that define the distinct formulations of the model, as well as the numerical results. These numerical results are connected with the analytical considerations presented in the Appendix. Finally, our conclusions are presented in section III.

% ###########################################################################

\section{Model and Results}

Our model is based on the Galam's majority-rule model \cite{galam_1999,galam_cont,krap_redner}. We consider a fully-connected population of $N=n_{A}+n_{B}$ agents with opinions $A$ or $B$ concerning a given subject. In this sense, we are considering a mean-field-like limit, since each agent can interact with all others. In this case, the microscopic dynamics disregards correlations, that will be taken into account after, when we will consider the model on regular lattices. The opinions are represented by Ising-like variables $o_{i}=\pm 1$ ($i=1,2,...,N$), and the initial concentration of each opinion is $0.5$ (disordered state). We will consider three distinct mechanisms in the formulation of our model, namely majority-rule dynamics, inflexibility and independence. Our objective is to analyze the critical behavior of the system, and in this case we will consider separately in the following subsections three distinct cases: (i) the majority-rule model with independent behavior, (ii) the majority-rule model with inflexible agents, and (iii) the majority-rule model with inflexible and independent individuals.

%%%%%%%%%%%%%%%%%%%%%%%%%%%%%%%%%%%%%%%

\subsection{Majority rule with Independence}

In this case, we consider that some individuals in the population can show a nonconformist behavior called independence \cite{sznajd_indep1,sznajd_indep2,sznajd_indep3,nuno_indep}. The following microscopic rules govern the dynamics:

\begin{enumerate}

\item A group of $3$ agents, say $(i,j,k)$, is randomly chosen;

\item With probability $q$ all the three agents in the group will act independently of the opinions of the group's individuals, i.e., independent of the majority/minority opinion inside the group. In this case, with probability $f$ all the three agents flip their opinions and with probability $1-f$ nothing occurs;

\item On the other hand, with probability $1-q$ the group follows the standard majority rule. In this case, all agents in the group follow the local majority opinion (if the opinion of one agent is different from the other two, the former flips alone).

\end{enumerate}

In the case where the 3 agents do not act independently, which occurs with probability $1-q$, the change of the states of the agents inside the group will occur according to the Galam's majority-rule model \cite{galam_1999,galam_cont,krap_redner}. The parameter $f$ can be related to the agents' flexibility \cite{sznajd_indep1}. As discussed in \cite{sznajd_indep2,sznajd_indep3,nuno_indep}, independence is a kind of nonconformity, and it acts on an opinion model as a kind of stochastic driving that can lead the model to undergo a phase transition. In fact, independence plays the role of a random noise similar to social temperature \cite{lalama}.

We analyze the critical behavior of the system, in analogy to magnetic spin systems, by computing the order parameter  
\begin{equation} \label{eq1}
O = \left\langle \frac{1}{N}\left|\sum_{i=1}^{N} o_{i}\right|\right\rangle ~, 
\end{equation}
where $\langle\, ...\, \rangle$ stands for time averages taken in the steady state. In addition to the time average, we have also considered configurational averages, i.e., averages over different realizations. The order parameter $O$ is sensitive to the unbalance between the two distinct opinions, and it plays the role of  the ``magnetization per spin'' in  magnetic systems. In addition, we also consider the fluctuations $\chi$ of the order parameter (or ``susceptibility'') 
\begin{equation} \label{eq2}
\chi =  N\,(\langle O^{2}\rangle - \langle O \rangle^{2})   
\end{equation}
and the Binder cumulant $U$, defined as \cite{binder}
\begin{equation} \label{eq3}
U   =   1 - \frac{\langle O^{4}\rangle}{3\,\langle O^{2}\rangle^{2}} \,.
\end{equation}

\begin{figure}[t]
\begin{center}
\vspace{3mm}
\includegraphics[width=0.55\textwidth,angle=0]{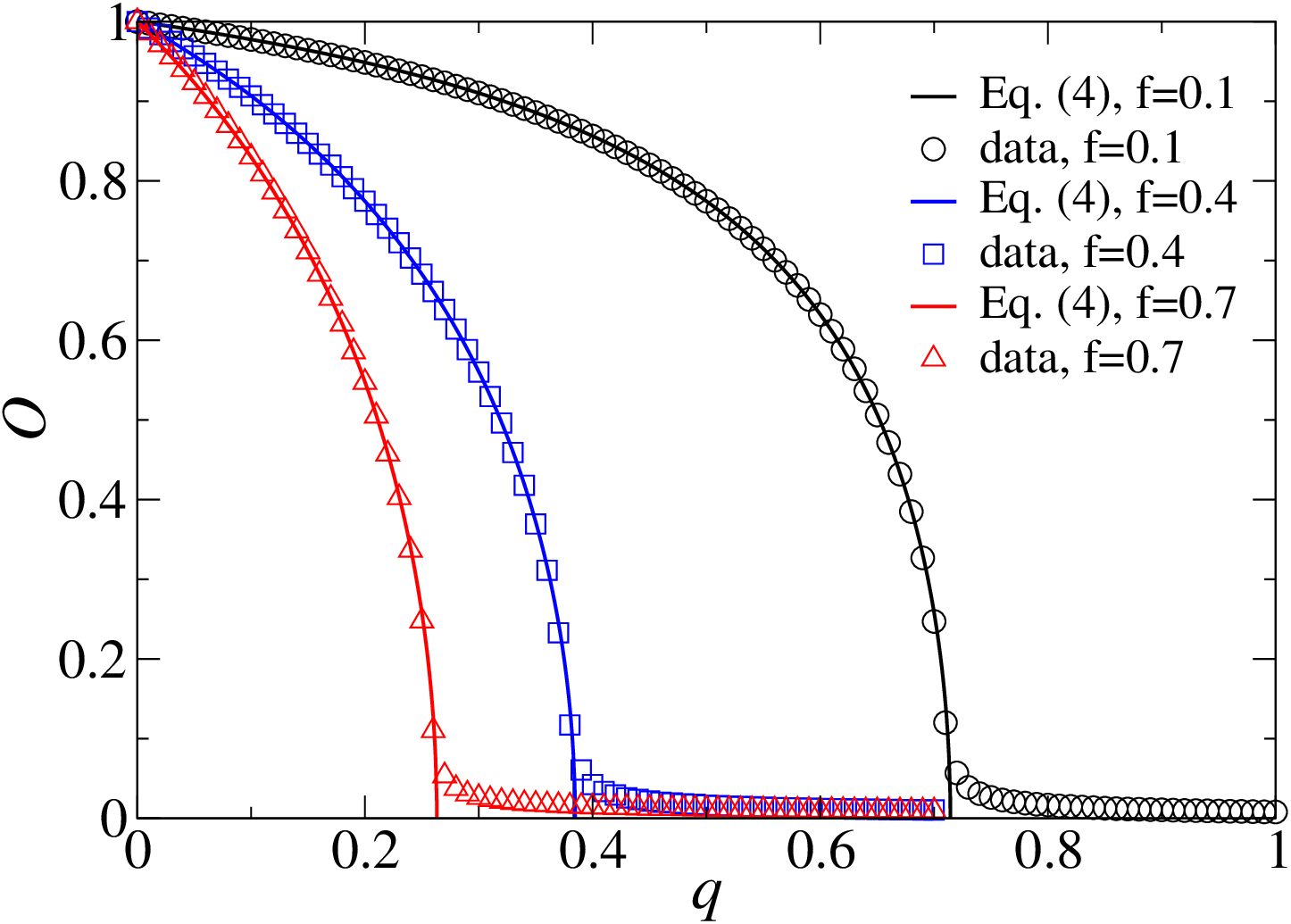}
\end{center}
\caption{(Color online) Order parameter $O$ versus the independence probability $q$ for typical values of the flexibility $f$, for the mean-field formulation of the model (no lattice). The symbols correspond to numerical simulations for population size $N=10000$ (averaged over $100$ simulations) and the full lines represent the analytical prediction, Eq. (\ref{eq4}).}
\label{fig1}
\end{figure}

\begin{figure}[t]
\begin{center}
\vspace{3mm}
\includegraphics[width=0.48\textwidth,angle=0]{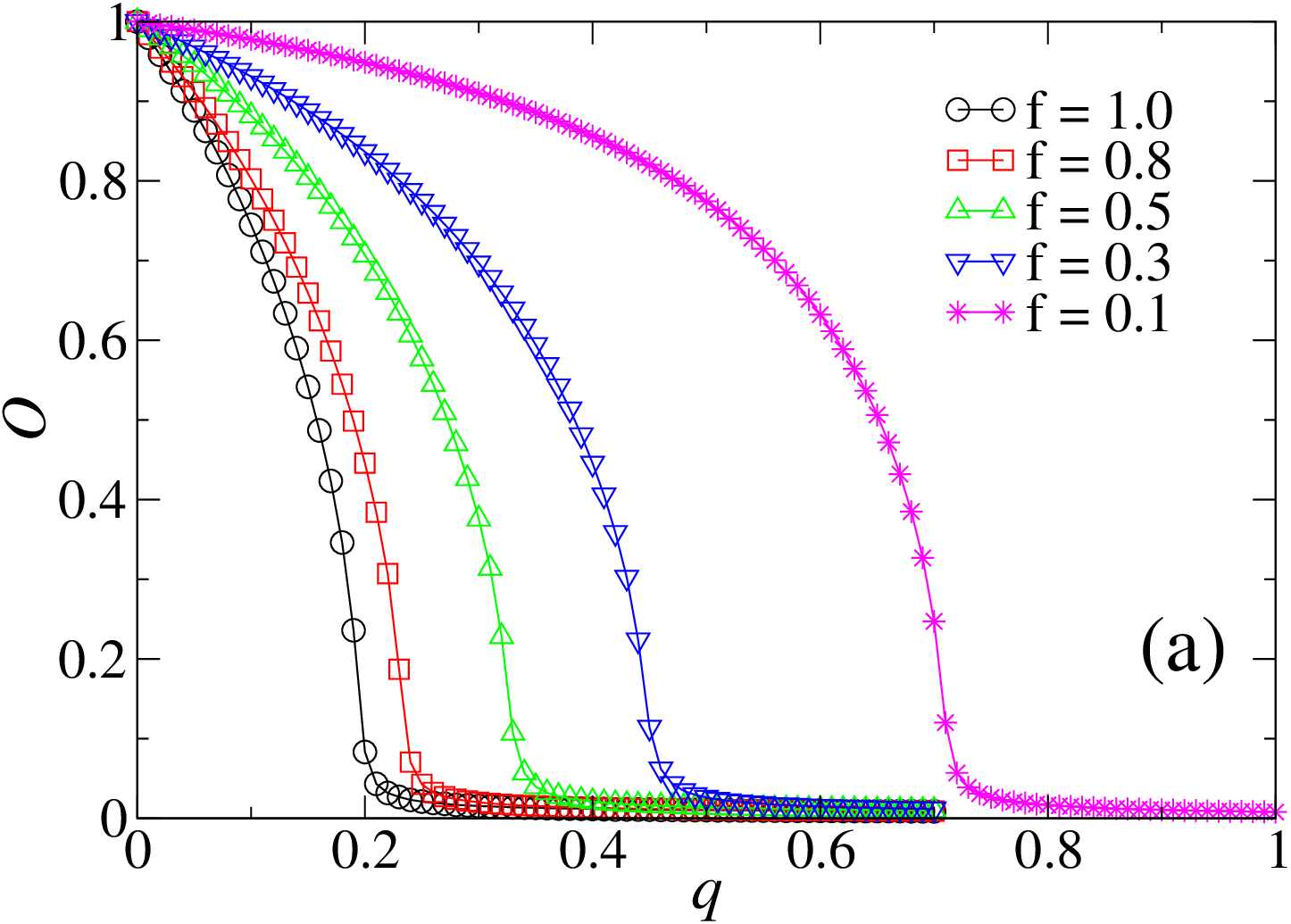}
\hspace{0.3cm}
\includegraphics[width=0.46\textwidth,angle=0]{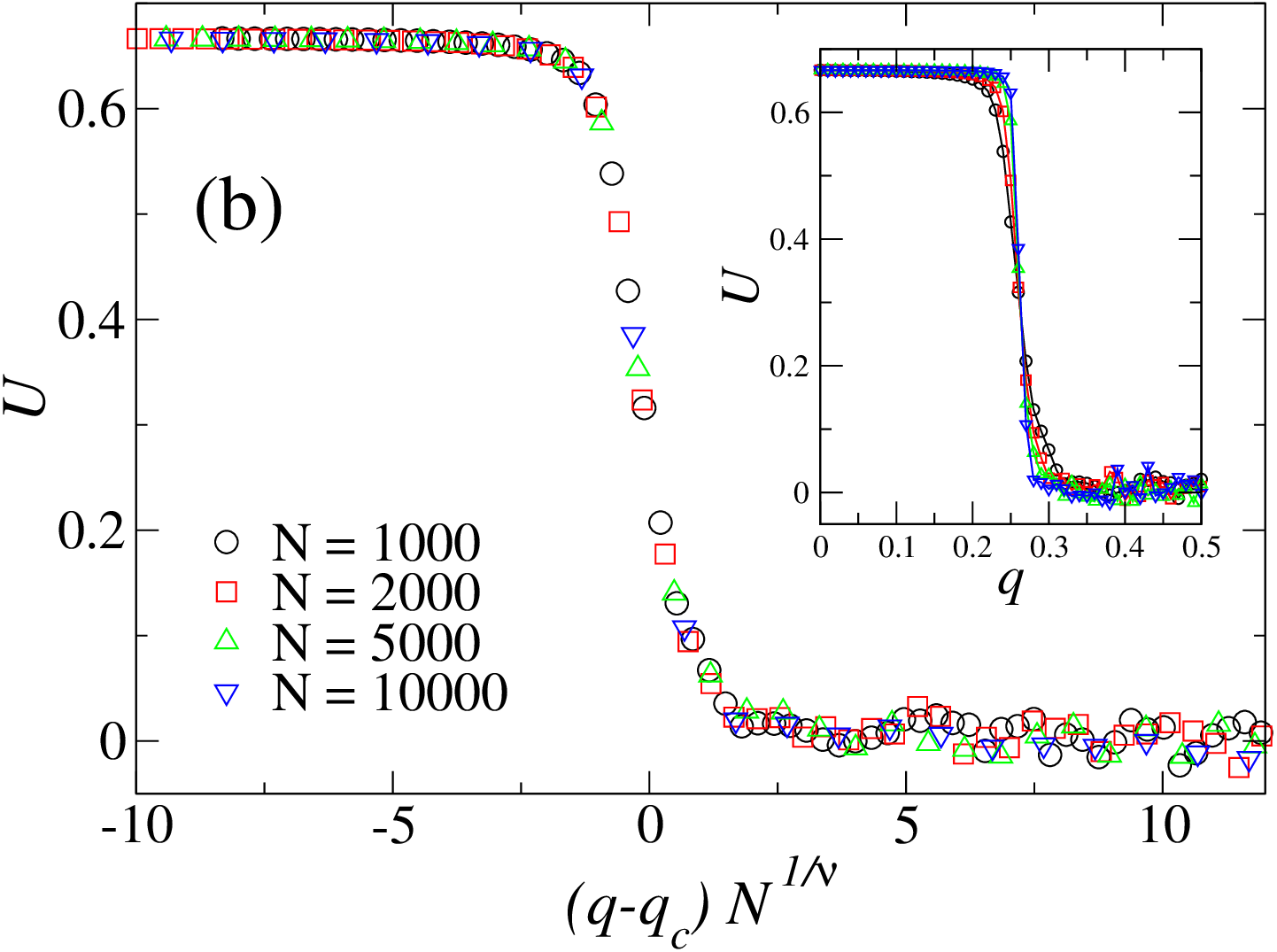}
\\
\vspace{1.0cm}
\includegraphics[width=0.46\textwidth,angle=0]{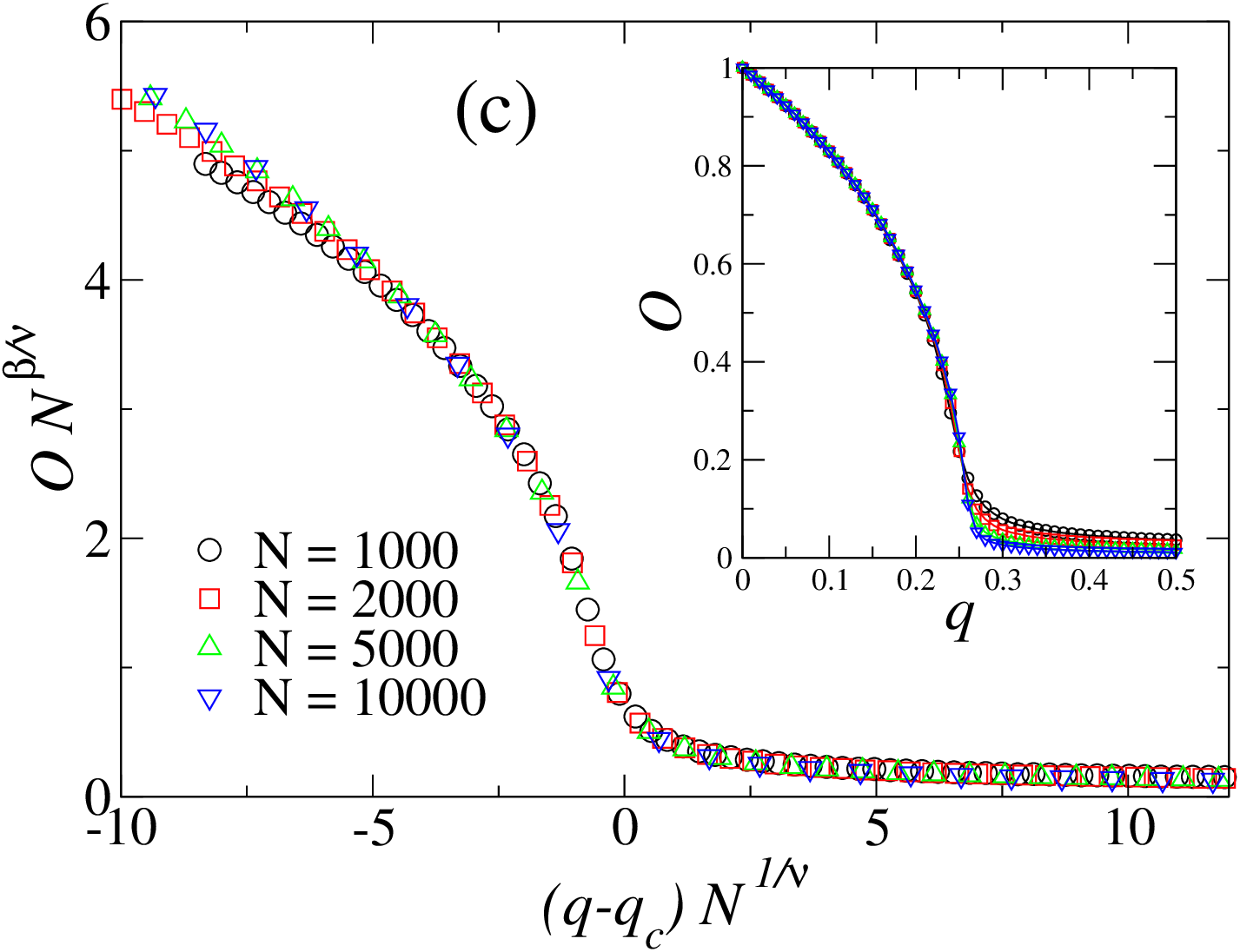}
\hspace{0.3cm}
\includegraphics[width=0.48\textwidth,angle=0]{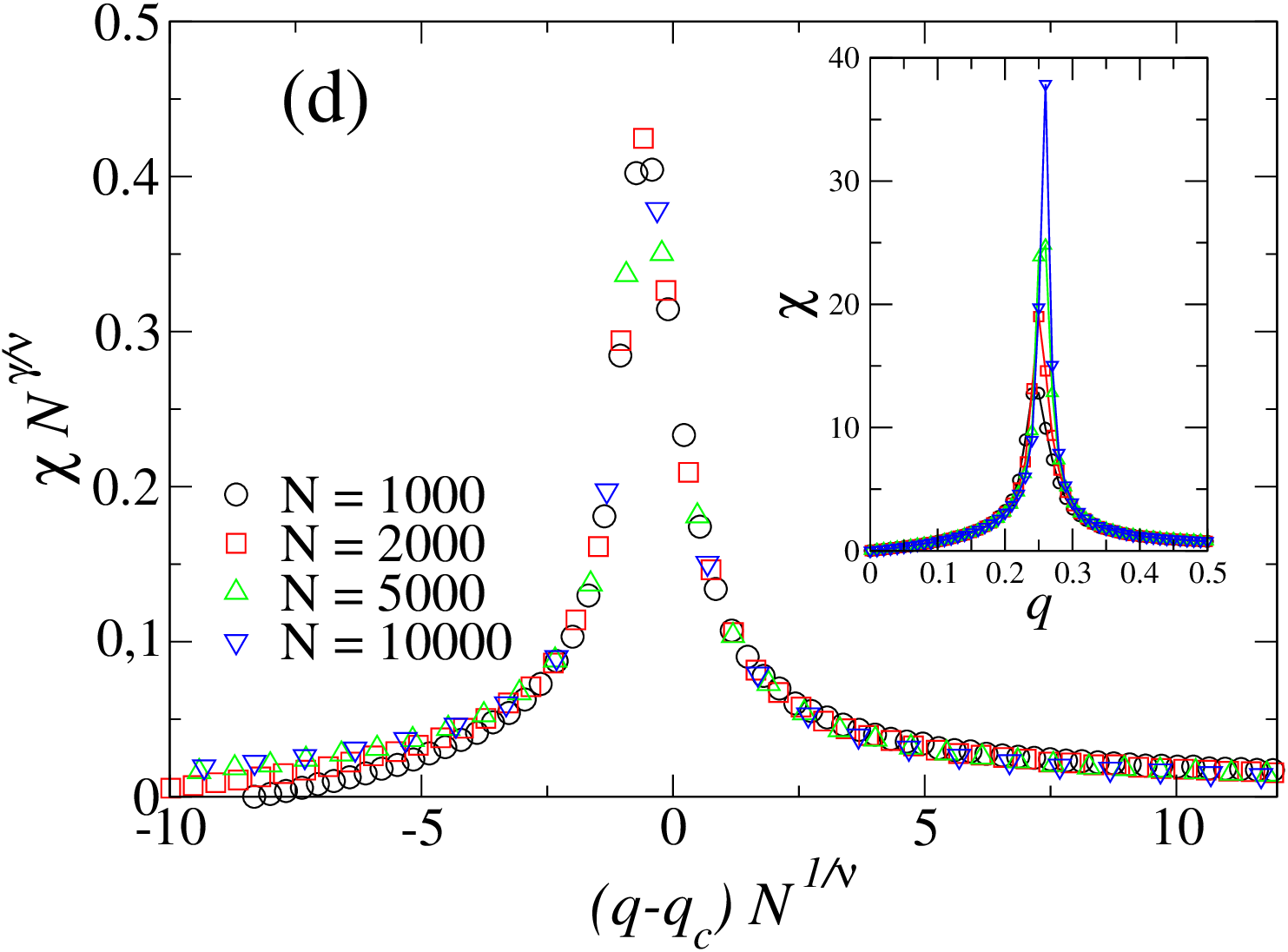}
\end{center}
\caption{(Color online) Numerical results for the mean-field formulation of the model with independence. (a) Order parameter $O$ versus $q$ for typical values of $f$ and population size $N=10000$. It is also exhibited the finite-size scaling analysis for $f=0.7$ (pannels b, c and d). We obtained $q_{c}\approx 0.264$, $\beta\approx 1/2$, $\gamma\approx 1$ and $\nu\approx 2$. Data are averaged over $100$ simulations.}
\label{fig2}
\end{figure}

As we are considering a mean-field formulation of the model, one can follow Refs. \cite{biswas,nuno_indep} to derive analytically the behavior of the stationary order parameter. The behavior of $O$ is given by (see Appendix 1)
\begin{equation} \label{eq4}
O = \left(1-\frac{4qf}{1-q}\right)^{1/2}
\end{equation}
or in the usual form $O\sim (q-q_{c})^{\beta}$, where
\begin{eqnarray} \label{eq5}
q_{c} = q_{c}(f) = \frac{1}{1+4\,f} ~,
\end{eqnarray}
\noindent
and we found a typical mean-field exponent $\beta=1/2$, as expected due to the mean-field character of the model. The comparison of Eq. (\ref{eq4}) with the numerical simulations of the model is given in Fig. \ref{fig1}, for typical values of the flexibility $f$. One can see an excellent agreement among the two results. Eq. (\ref{eq5}) also predicts that there is an order-disorder transition for all values of $f>0$, which was confirmed numerically, see Fig. \ref{fig2} (a).

%%%%%%%%%%%%%%%%%%%%%%%%%%%%%%%%%%%%%%%%%%%%%%%%%%%%%%%%%%%%%%%%%%%%%%%%%%
%%%%%%%%%%%%%%%%%%%%%%%%%%%%%%%%%%%%%%%%%%%%%%%%%%%%%%%%%%%%%%%%%%%%%%%%%%%

%%%%%%%%%%%%%%%%%%%%%%%%%%%%%%%%%%%%%%%%%%%%%%%%%%%%%%%%%%%%%%%%%%%%%%%%%%
%%%%%%%%%%%%%%%%%%%%%%%%%%%%%%%%%%%%%%%%%%%%%%%%%%%%%%%%%%%%%%%%%%%%%%%%%%%

We also estimated the critical exponents for many values of $f$. As a typical example, we exhibit in Fig. \ref{fig2} the finite-size scaling (FSS) analysis for $f=0.7$ (see pannels b, c and d). The critical values $q_{c}$ were identified by the crossing of the Binder cumulant curves, as can be seen in the inset of Fig. \ref{fig2} (b), and the critical exponents $\beta$, $\gamma$ and $\nu$ were found by the best collapse of data. For all values of $f$ we found $\beta\approx 1/2$, $\gamma\approx 1$ and $\nu\approx 2$, which suggests a universality of the order-disorder phase transition. In particular, the numerical estimates of the exponent $\beta$ agree with Eq. (\ref{eq4}), that predicts $\beta=1/2$ for all values of $f$. Notice that the exponents $\beta$ and $\gamma$ are typical Ising mean-field exponents, which is not the case for $\nu$. This same discrepancy was observed in other discrete opinion models \cite{biswas,nuno_celia_victor,nuno_indep}, and was associated with a superior critical dimension $D_{c}=4$, that leads to an effective exponent $\nu^{'}=1/2$, obtained from $\nu=D_{c}\,\nu^{'}=2$. In this case, one can say that our model is in the same universality class of the kinetic exchange opinion models with two-agent interactions \cite{biswas,biswas2,nuno_indep,nuno_celia_victor}, as well as in the mean-field Ising universality class.

%%%%%%%%%%%%%%%%%%%%%%%%%%%%%%%%%%%%%%%%%%%%%%%%%%%%%%%%%%%%%%%%%%%%%%%%%%
\begin{figure}[t]
\begin{center}
\vspace{3mm}
\includegraphics[width=0.48\textwidth,angle=0]{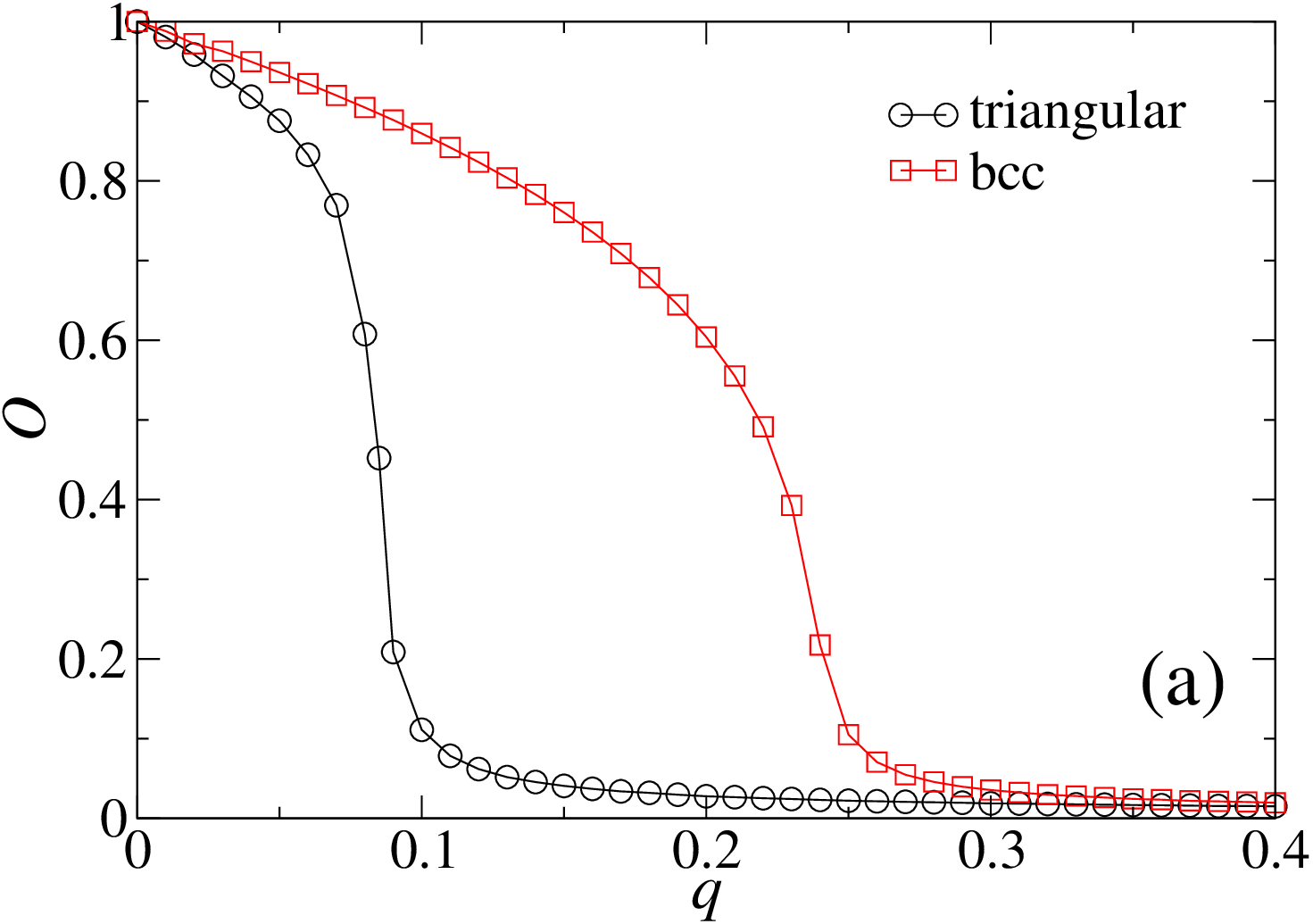}
\hspace{0.3cm}
\includegraphics[width=0.48\textwidth,angle=0]{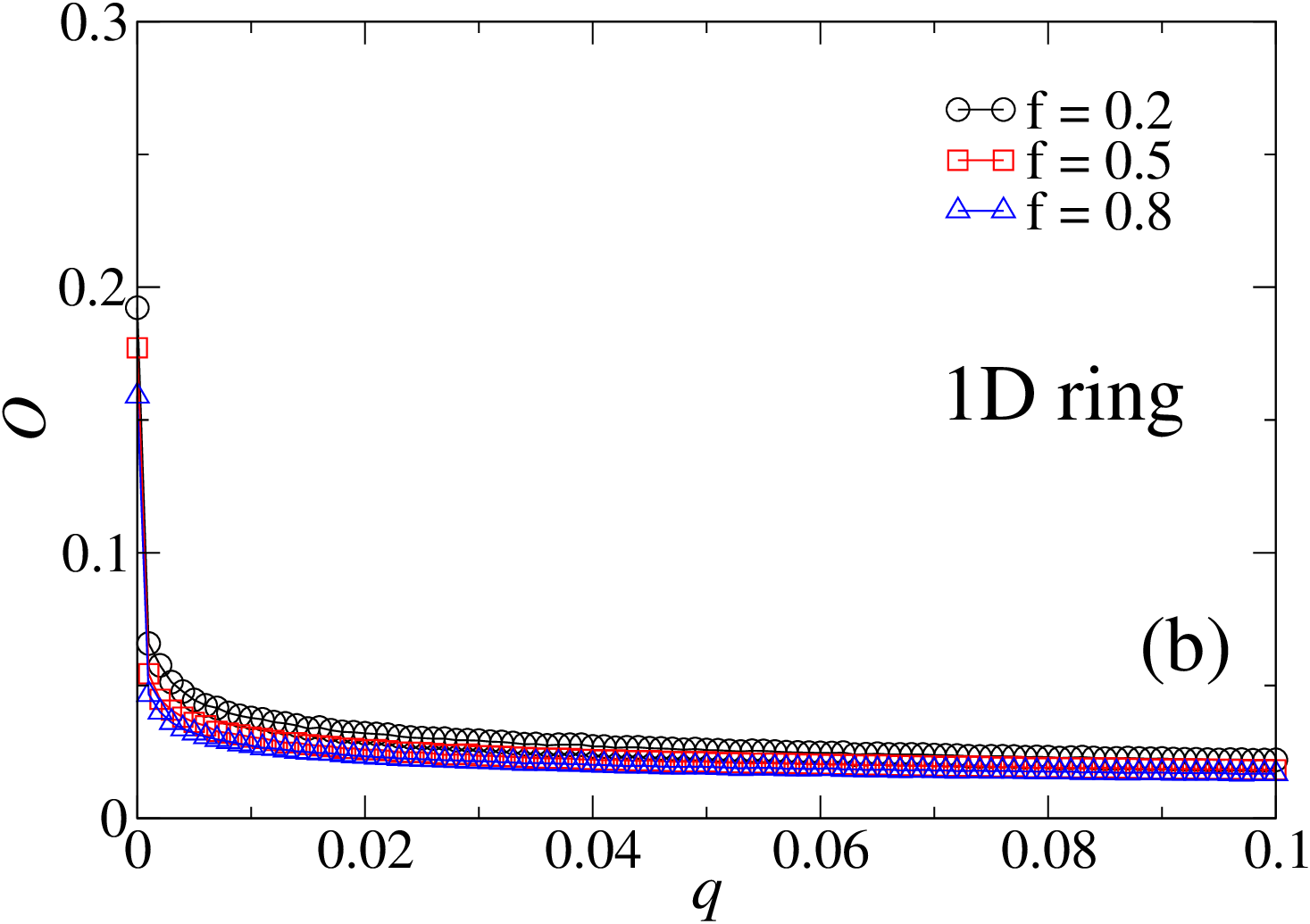}
\end{center}
\caption{(Color online) Order parameter $O$ versus the independence probability $q$ for the 2D (triangular lattice, $L=100$) and 3D (bcc lattice, $L=20$) cases, considering $f=0.5$ (a). One can see the typical behavior of a phase transition. We also shown $O$ versus $q$ for the model defined on a 1D ring with $N=10000$ sites (b). In this case, the results for distinct values of $f$ suggest the absence of a phase transition. All data are averaged over $100$ simulations.}
\label{fig3_new}
\end{figure}
%%%%%%%%%%%%%%%%%%%%%%%%%%%%%%%%%%%%%%%%%%%%%%%%%%%%%%%%%%%%%%%%%%%%%%%%%%%

To test the universality of the model under the presence of a topology, we simulated the dynamics on two distinct lattices, namely a two-dimensional triangular lattice and a three-dimensional body-centered cubic (bcc) lattice. In this case, the presence of a topology will introduce correlations in the system, and we expected that the mean-field results are not valid anymore. The lattices were built as folows. The triangular lattice was built from a finite $L\times L$ square lattice with extra bonds along one diagonal direction. In this case, each group of 3 agents is chosen as follows. First, we choose an agent at random, say $i$. Then, we choose at random two nearest neighbors of $i$ (say $j$ and $k$), in a way that each one of the 3 agents ($i$, $j$ and $k$) is a neighbor of the other two agents, forming a triangle. On the other hand, the bcc lattice was built from a cubic structure with linear size $2L$, and each group contains 5 agents that were chosen as follows. First, we choose a random plaquette of 4 neighbor sites, forming a square. The fifth site is randomly chosen between the 2 possible sites in order to form a pyramid. A typical behavior of the order parameter as a function of $q$ is shown in Fig. \ref{fig3_new} (a) for both cases (2D and 3D, considering $f=0.5$), where one can observe a typical behavior of an order-disorder transition. Considering distinct values of $f$, we performed a FSS analysis ir order to estimate the critical exponents (not shown). Thus, for the 2D lattice we obtained the same critical exponents of the 2D Ising model for all values of $f$, i.e., $\beta\approx 0.125$, $\nu\approx 1.0$ and $\gamma\approx 1.75$, and for the 3D lattice we obtained the same critical exponents of the 3D Ising model for all values of $f$, i.e., $\beta\approx 0.32$, $\nu\approx 0.63$ and $\gamma\approx 1.24$ \cite{gould}. These results suggest that considering a bidimensional (tridimensional) system the model is in the universality class of the 2D (3D) Ising model. Finally, we simulated the model on an one-dimensional ring, where each 3-agents group was formed by a randomly chosen site and its two nearest neighbors. Typical results for the order parameter as a function of $q$ are exhibited in Fig. \ref{fig3_new} (b). In this case, the results suggest that there is no order-disorder transition, as in the 1D Ising model. In this case, considering the results for 1D, 2D and 3D lattices and also for the mean-field case, one can say that the majority-rule model with independent behavior is in the Ising model universality class.

\begin{figure}[t]
\begin{center}
\vspace{3mm}
\includegraphics[width=0.55\textwidth,angle=0]{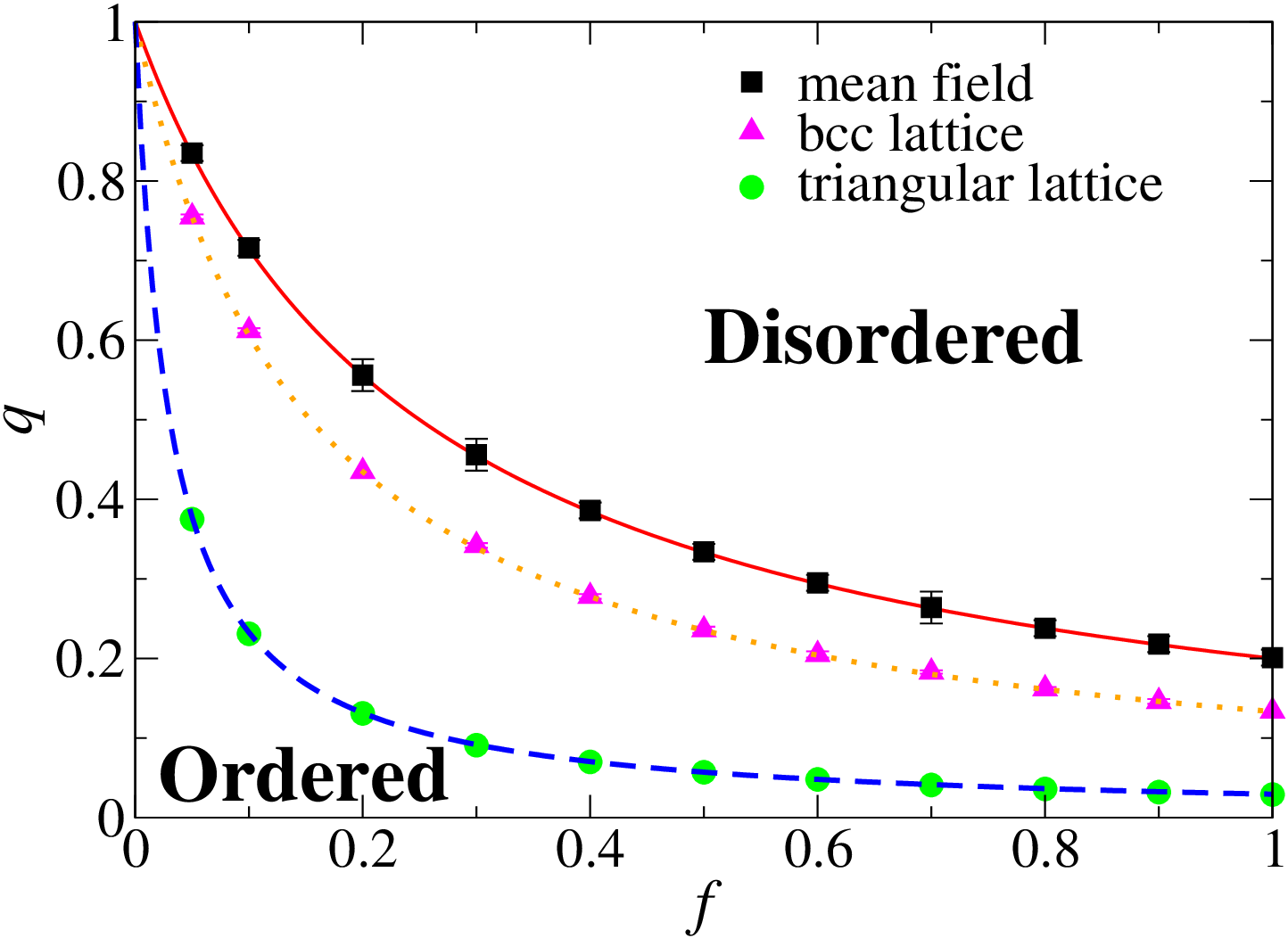}
\end{center}
\caption{(Color online) Comparative phase diagram of the majority-rule model with independent behavior of the agents, for the mean-field, triangular lattice and bcc lattice cases, separating the Ordered and the Disordered phases. The symbols are the numerical estimates of the critical points, obtained from the crossing of the Binder cumulant curves for different population sizes. The lines are given by the analytical form, Eq. (\ref{eq6}), with $a=4$ for the mean-field case [full line, according to Eq. (\ref{eq5})], $a\approx 33$ (dashed line) for the triangular lattice and $a\approx 6.5$ (dotted line) for the bcc lattice, the last two values obtained from data fits.}
\label{fig3}
\end{figure}

The comparative phase diagram (mean field x 2D x 3D) is exhibited in Fig. \ref{fig3}, where we plot for the fully-connected case the analytical solution for $q_{c}$, Eq. (\ref{eq5}). The behavior is qualitatively similar in all cases, suggesting a frontier of the form 
\begin{equation} \label{eq6}
q_{c} = \frac{1}{1+a\,f} ~,
\end{equation}
where $a=4$ for the mean-field case, $a\approx 33$ for the triangular lattice and $a\approx 6.5$ for the bcc lattice (these last two values were obtained by a fit of the data exhibited in Fig. \ref{fig3}). 

%%%%%%%%%%%%%%%%%%%%%%%%%%%%%%%%%%%%%%%%%%%%%%%%%%%%%%%%%%%%%%%%%%%%%%%%%%
%%%%%%%%%%%%%%%%%%%%%%%%%%%%%%%%%%%%%%%%%%%%%%%%%%%%%%%%%%%%%%%%%%%%%%%%%%%

Notice that the mean-field analytical calculation, Eq. (\ref{eq5}), overestimates the critical points $q_{c}$, as it is common in mean-field approximations. However, our calculations predict the occurrence of order-disorder phase transitions, as well as correctly predicts the form $q_{c}=1/(1+a\,f)$ between $q_{c}$ and $f$.

From the phase diagram of this formulation of the model we can see that the increase of the flexibility parameter $f$ leads to the decrease of $q_{c}$, as also indicated in Eq. (\ref{eq5}). This can be understood as follows. The increase of $f$ leads the agents to perform more independent opinion changes or spin flips (which represents a nonconservative society). This action tends to disorder the system even for a small value of the independence probability $q$, which decrease the critical point $q_{c}$.

Notice that we obtained here for the mean-field case the same result for $q_{c}$ obtained in ref. \cite{sznajd_indep1}, where the independent behavior was considered in the Sznajd model. Indeed, in the mean-field formulation of the Sznajd model, the dynamics is very similar to the mean-field majority-rule dynamics for groups of size $3$, which explains the identical result. However, in the mentioned reference, the model was not mapped in any universality class.

%%%%%%%%%%%%%%%%%%%%%%%%%%%%%%%%%%%%%%%

\subsection{Majority rule with Inflexibility}

As a second formulation of our model, we consider the majority-rule dynamics with the presence of some agents with the inflexibility characteristic, individuals whose stubbornness makes them reluctant to change their opinions \cite{moscovici,galam_inflex,schneider,jiang,martins,mobilia,galam2011,nuno_celia_victor}. As in \cite{nuno_celia_victor}, we have considered a fraction $d$ of agents that are averse to change their opinions. The following microscopic rules govern the dynamics:

\begin{enumerate}

\item A group of $3$ agents, say $(i,j,k)$, is randomly chosen;

\item We verify if there is a majority of 2 (say $i$ and $j$) in favor of a given opinion $A$ or $B$, and in this case the other (say $k$) is a supporter of the minority opinion;

\item If agent $k$ is a flexible individual, he/she will follow the local majority and flip his state $o_{k}\to-o_{k}$, otherwise nothing occurs.

\end{enumerate}

In this case, the frozen states of the inflexible agents work in the model as the introduction of a quenched disorder. As in magnetic systems \cite{spin_glass}, one can expect that a disorder can induce/suppress a phase transition, as was also observed in the kinetic exchange opinion model with the presence of inflexibles \cite{nuno_celia_victor}.

%%%%%%%%%%%%%%%%%%%%%%%%%%%%%%%%%%%%%%%%%%%%%%%%%%%%%%%%%%%%%%%%%%%%%%%%%%
\begin{figure}[t]
\begin{center}
\vspace{3mm}
\includegraphics[width=0.55\textwidth,angle=0]{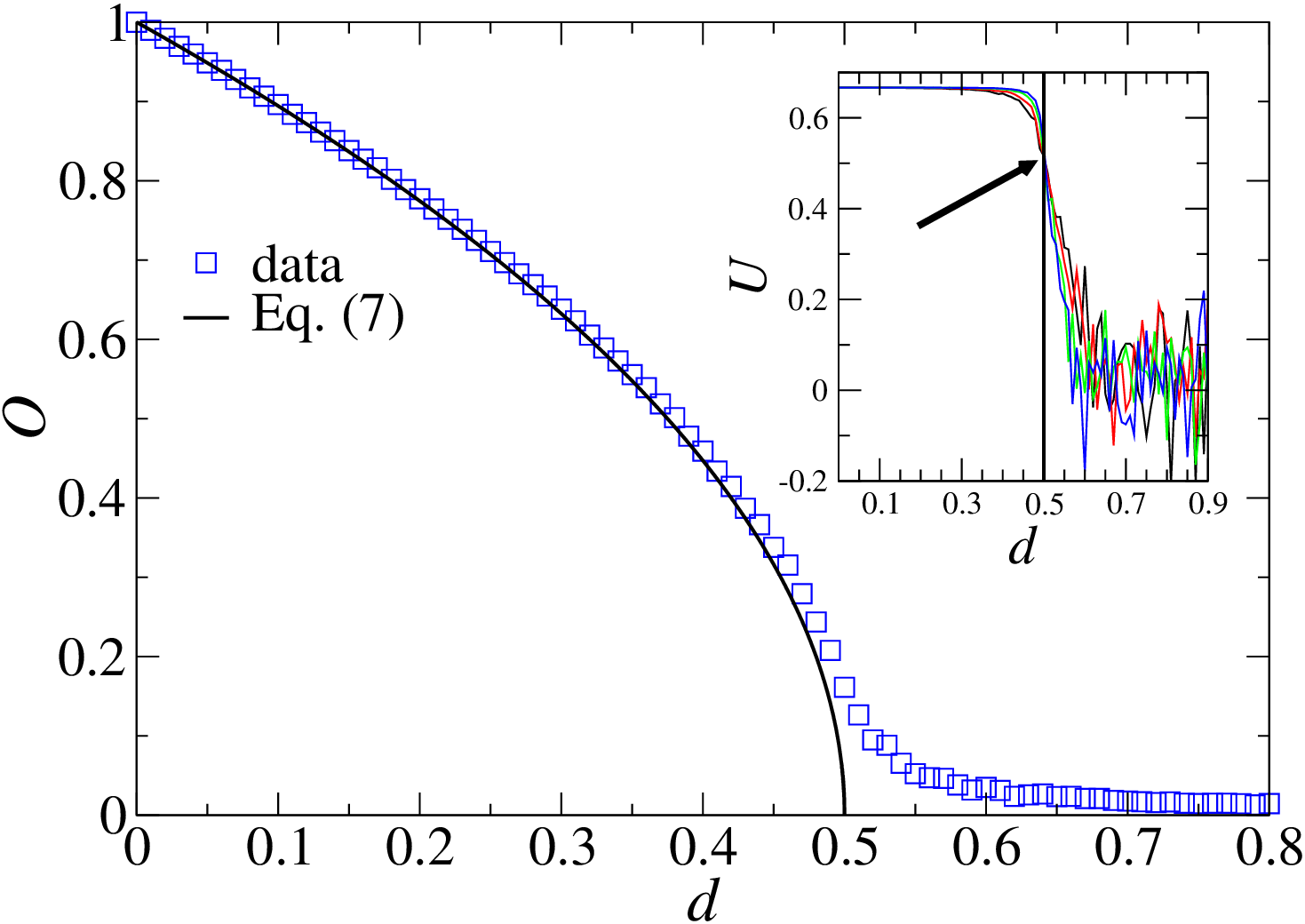}
\end{center}
\caption{(Color online) Order parameter $O$ versus the fraction $d$ of inflexible individuals for the mean-field formulation of the model (main plot). The squares are the numerical results for population size $N=10000$ and the full line is the analytical prediction, Eq. (\ref{eq7}). It is also exhibited in the inset the Binder cumulant curves for different sizes, showing the crossing of the curves for $d_{c}\approx 0.5$, in agreement with Eqs. (\ref{eq7}) and (\ref{eq8}). Data are averaged over $100$ simulations.}
\label{fig4}
\end{figure}
%%%%%%%%%%%%%%%%%%%%%%%%%%%%%%%%%%%%%%%%%%%%%%%%%%%%%%%%%%%%%%%%%%%%%%%%%%%

As in the previous case (subsection A), one can derive analytically the behavior of the order parameter in this mean-field formulation of the model. The dependence of the order parameter with the the fraction $d$ of inflexibles is given by (see Appendix 2)
\begin{equation} \label{eq7}
O = (1-2\,d)^{1/2} ~,
\end{equation}
\noindent
or in the usual form $O\sim (d-d_{c})^{\beta}$, where
\begin{eqnarray} \label{eq8}
d_{c} = \frac{1}{2} ~,
\end{eqnarray}
and again we found a typical mean-field exponent $\beta=1/2$, as expected due to the mean-field character of the model. The comparison of Eq. (\ref{eq7}) with the numerical simulations of the model is given in Fig. \ref{fig4}. In addition, we also show in the inset of Fig. \ref{fig4} the Binder cumulant curves for different population sizes, where one can observe the crossing of the curves at $d_{c}\approx 0.5$, in agreement with Eqs. (\ref{eq7}) and (\ref{eq8}). Furthermore, a complete FSS analysis (not shown) give us $\beta\approx 1/2$, $\gamma\approx 1$ and $\nu\approx 2$, i.e., the same values obtained for the model presented in the subsection A. Thus, the presence of intransigents in the population leads the system to an order-disorder transition at a critical density $d_{c}=1/2$, and this transition is in the mean-field Ising model universality class.

%%%%%%%%%%%%%%%%%%%%%%%%%%%%%%%%%%%%%%%%%%%%%%%%%%%%%%%%%%%%%%%%%%%%%%%%%%
\begin{figure}[t]
\begin{center}
\vspace{3mm}
\includegraphics[width=0.55\textwidth,angle=0]{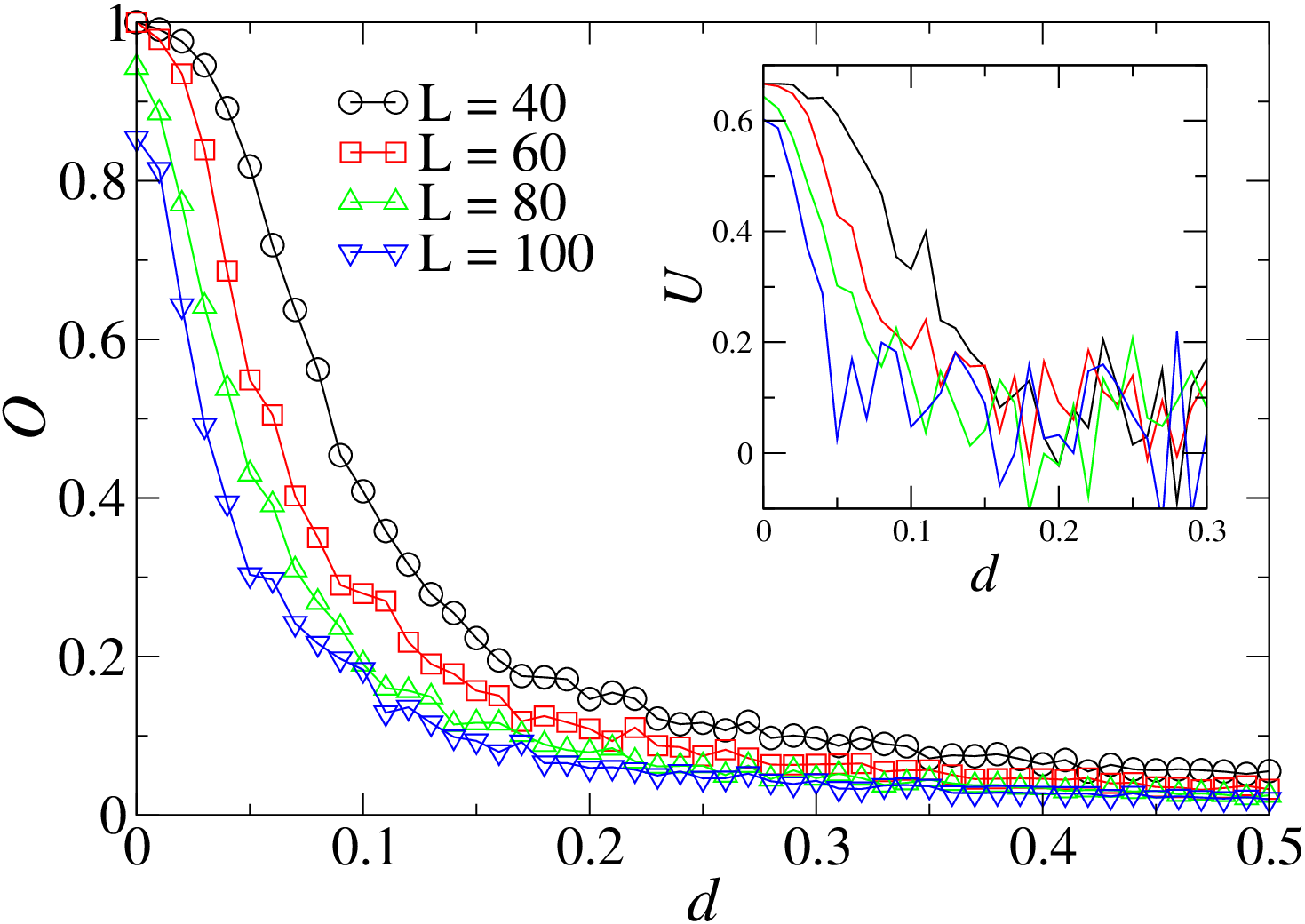}
\end{center}
\caption{(Color online) Order parameter $O$ versus the fraction $d$ of inflexible individuals for the model with no independence defined on triangular lattices of distinct sizes $L$ (main plot). It is also exhibited in the inset the Binder cumulant curves for different sizes, showing no crossing of the curves. Data are averaged over $100$ simulations.}
\label{fig5}
\end{figure}
%%%%%%%%%%%%%%%%%%%%%%%%%%%%%%%%%%%%%%%%%%%%%%%%%%%%%%%%%%%%%%%%%%%%%%%%%%%

As a final observation of this subsection, we also simulated (as in the previous section) the majority-rule model with inflexible agents on a two-dimensional triangular lattice, in order to test the universality of the model in comparison with the Ising model. The results are exhibited in Fig. \ref{fig5}. One can see that the order parameter $O$ (at least for the larger sizes) does not present the typical behavior of a phase transition, i.e., the usual change of concavity of the curves. In addition, in the inset of Fig. \ref{fig5} one can see that the Binder cumulant curves do not cross. A similar behavior was also reported in \cite{nuno_celia_victor} for another opinion model with inflexibility. In this case, those results suggest that the inclusion of inflexibility as we done here works in the model as a quenched disorder, and it destroys the phase transition in small dimensions like $D=2$. In order to verify this hypothesis, we simulated the model on square lattices. In such case, we randomly choose a lattice site, and the group is formed by this random individual and his/her four nearest neighbors, forming a group of size 5 as in \cite{chen_redner}. The behavior of $O$ and $U$ are very similar to the ones observed for the triangular lattice (not shown), suggesting that there is no order-disorder transition for $D=2$. In some magnetic models such type of destruction due to quenched disorder was also observed \cite{spin_glass,jensen,mercaldo}.

%%%%%%%%%%%%%%%%%%%%%%%%%%%%%%%%%%%%%%%

\subsection{Majority rule with Independence and Inflexibility}

As a third formulation of our model, we consider the majority-rule dynamics where agents can exhibit the independent behavior, as well as inflexibility. In this case, the model carries the rules of the two previous models (subsections A and B), namely:

\begin{enumerate}

\item A group of $3$ agents, say $(i,j,k)$, is randomly chosen;

\item With probability $q$ each one of the three agents in the group will act independently of the opinions of the group's individuals, provided he/she is not an inflexible individual. Thus, with probability $f$ all flexible agents flip their opinions and with probability $1-f$ nothing occurs;

\item On the other hand, with probability $1-q$ the group follows the standard majority rule. In this case, each flexible agent follows the local majority opinion.

\end{enumerate}

Notice that, even if the agents decide to act independently of the group's opinions, we will not see necessarily 3 changes of opinions, as in the model of subsection B. Indeed, the 3 agents can change their opinions, but we can have two, one or even zero spin flips due to the frozen states of the inflexible agents.

%%%%%%%%%%%%%%%%%%%%%%%%%%%%%%%%%%%%%%%%%%%%%%%%%%%%%%%%%%%%%%%%%%%%%%%%%%
\begin{figure}[t]
\begin{center}
\vspace{3mm}
\includegraphics[width=0.55\textwidth,angle=0]{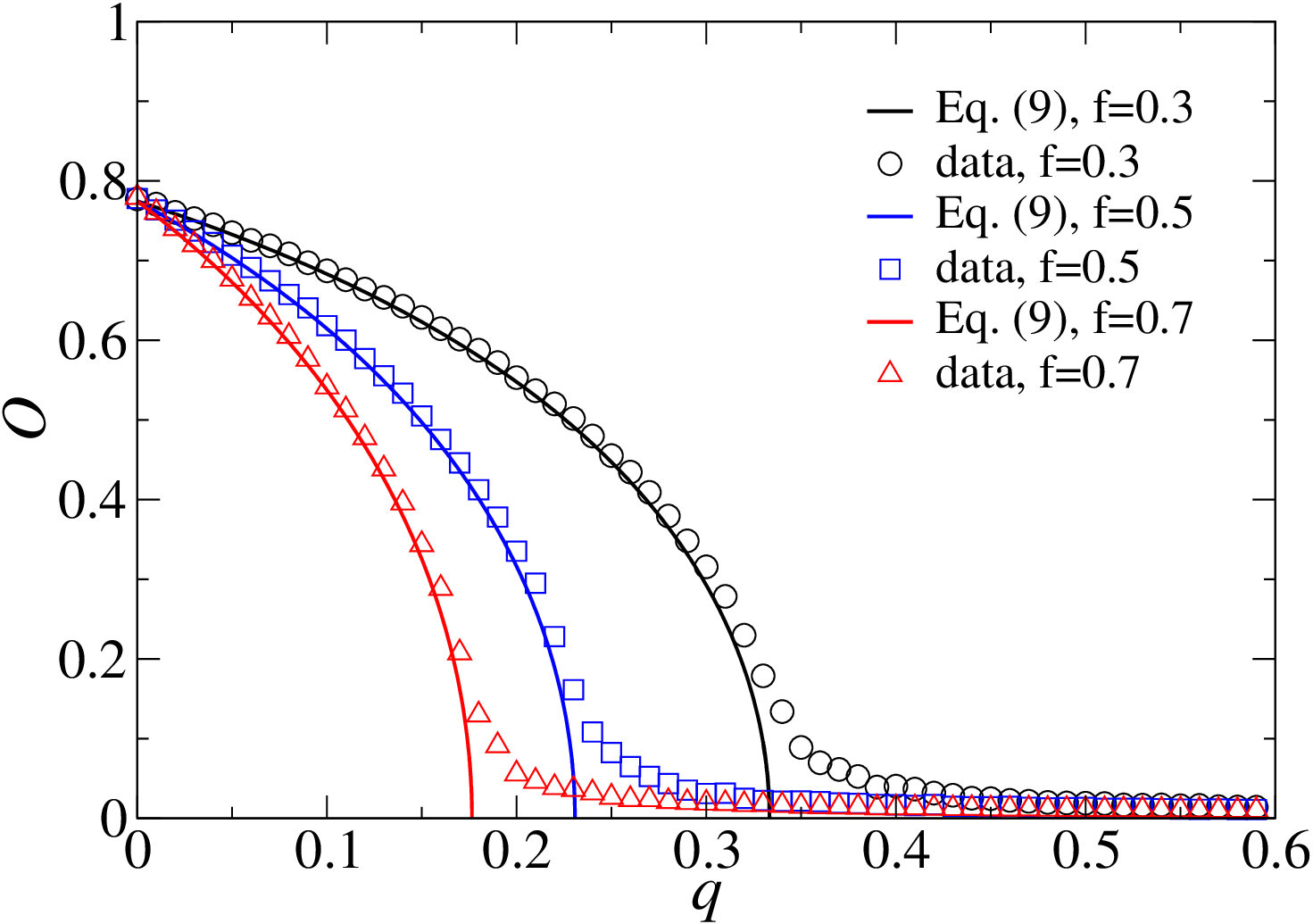}
\end{center}
\caption{(Color online) Order parameter $O$ versus the independence probability $q$ for $d=0.2$ and typical values of the flexibility $f$, for the mean-field formulation of the model. The symbols correspond to numerical simulations for population size $N=10000$ (averaged over $100$ simulations) and the full lines represent the analytical prediction, Eq. (\ref{eq9}).}
\label{fig6}
\end{figure}
%%%%%%%%%%%%%%%%%%%%%%%%%%%%%%%%%%%%%%%%%%%%%%%%%%%%%%%%%%%%%%%%%%%%%%%%%%%

\begin{figure}[t]
\begin{center}
\vspace{3mm}
\includegraphics[width=0.55\textwidth,angle=0]{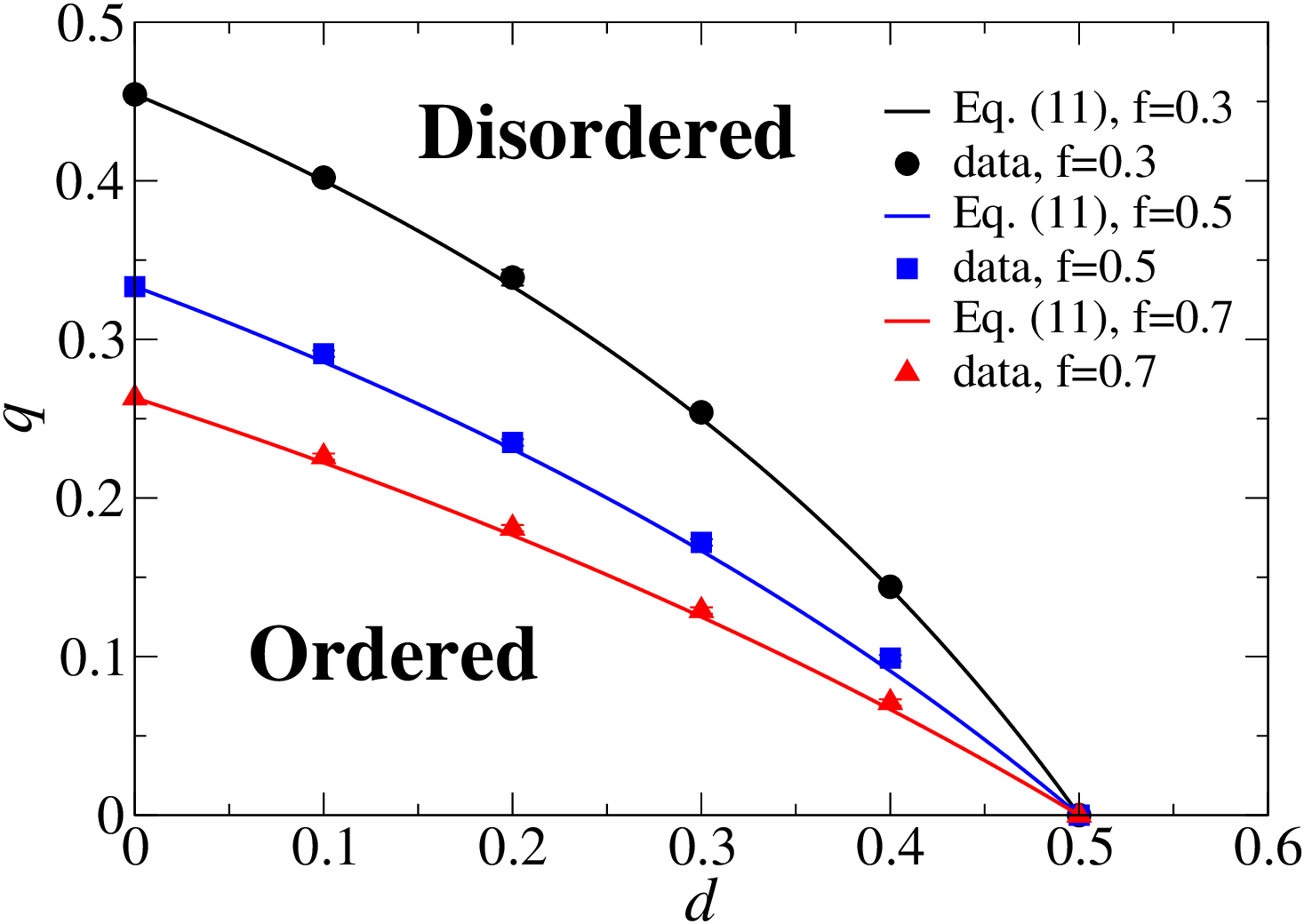}
\end{center}
\caption{(Color online) Phase diagram of the mean-field model with independence and inflexibility, in the plane $q$ versus $d$ for typical values of $f$. The symbols are the numerical estimates of the critical points, obtained from the crossing of the Binder cumulant curves for different population sizes. The full lines are the analytical form given by Eq. (\ref{eq11}).}
\label{fig7}
\end{figure}

As in the previous cases, one can derive analytically the behavior of the order parameter as a function of the fraction $d$ of inflexibles and the independence probability $q$, in the mean-field formulation of the model. The calculations give us (see Appendix 3)
\begin{equation} \label{eq9}
O = [(1-d)^{2}\,(1-\theta)]^{1/2} ~,
\end{equation}
\noindent
where $\theta=\theta(q,d,f)$ is given by
\begin{equation} \label{eq10}
\theta=\frac{4}{(1-q)\,(1-d)^{2}}\left\{\,f\,q+(1-q)\,\frac{d^{2}}{4}\right\} ~.
\end{equation}
Writing the order parameter in the the usual form $O\sim (q-q_{c})^{\beta}$, one obtains
\begin{eqnarray} \label{eq11}
q_{c} = q_{c}(f,d) = \frac{1-2\,d}{1+4\,f-2\,d}
\end{eqnarray}
and again we found a typical mean-field exponent $\beta=1/2$, as expected due to the mean-field character of the model. Notice that we recover the results of Eqs. (\ref{eq5}) and (\ref{eq8}) for $d=0$ (no inflexibility) and $q_{c}=0$ (no independence), respectively. The comparison of Eq. (\ref{eq9}) with the numerical simulations of the model for $d=0.2$ and typical values of $f$ is given in Fig. \ref{fig6}. In addition, a complete FSS analysis (not shown) for many values of $f$ give us $\beta\approx 1/2$, $\gamma\approx 1$ and $\nu\approx 2$, i.e., the same values obtained for the model presented in the previous subsections. Thus, this formulation of the model also leads the system to undergoes phase transitions in the same universality class of the mean-field Ising model. We also obtained from the FSS analysis the critical points $q_{c}$ for typical values of $d$ and $f$. The comparison among the numerical estimates and the analytical prediction of Eq. (\ref{eq11}) is shown in the phase diagram of Fig. \ref{fig7}.

%%%%%%%%%%%%%%%%%%%%%%%%%%%%%%%%%%%%%%%%%%%%%%%%%%%%%%%%%%%%%%%%%%%%%%%%%%
%%%%%%%%%%%%%%%%%%%%%%%%%%%%%%%%%%%%%%%%%%%%%%%%%%%%%%%%%%%%%%%%%%%%%%%%%%%

From the phase diagram of this formulation of the model we can see that the decrease of the flexibility parameter $f$, related to the independent behavior, makes the ordered phase greater, for a given value of $d<d_{c}=1/2$. As in the case with no inflexibility, the increase of $f$ leads the agents to perform more independent spin flips, and this action tends to disorder the system even for a small value of the independence probability $q$, which decrease the critical point $q_{c}$. The presence of intransigent agents reinforces this behavior, leading to the decrease of the ordered phase for increasing values of $d$.

% ############################################################################

\section{Final remarks}   

In this work, we have studied a discrete-state opinion model where each agent carries one of two possible opinions, $\pm 1$. For this purpose, we considered three distinct mechanisms to model the social behavior of the agents: majority-rule dynamics, inflexibility and independence. Our target was to study the critical behavior of the opinion model under the presence of the mentioned mechanisms. Thus, we performed computer simulations of the model, and some analytical calculations complemented the numerical analysis. Let us remember that the original majority-rule model presents only absorbing consensus states with all opinions $+1$ or $-1$  \cite{galam_1999,galam_cont,krap_redner}. 

First we considered the majority-rule model with independence in a fully-connected population. In this case, there is a probability $q$ that the 3 agents forming a group behave as independent individuals, changing opinion with probability $f$ and keeping opinion with probability $1-f$. This mechanism acts in the system as a social temperature. In this case, we showed that independence induces an order-disorder transition for all values of $f>0$, with the critical points $q_{c}$ being a function of $f$. In addition, the model is in the same universality class of the mean-field Ising model and of the kinetic exchange opinion models. In the ordered phase there is a coexistence of both opinions $\pm 1$, but one of them is a majority in the population. We observed that the larger the flexibility $f$ (nonconservative societies), the smaller the value of $q$ needed to disorder the system. In other words, for a population debating a subject with two distinct choices, it is easier to reach a final decision for a small flexibility concerning the independent behavior, as observed in conservative societies. Consensus states were obtained only for $q=0$ or $f=0$. As a test to the universality of the model, we simulated it on triangular and on bcc lattices, and we found the same critical exponents of the 2D and the 3D Ising models, respectively. In addition, simulations on 1D lattices suggest the absence of a phase transition. All those results suggest the majority-rule model with independence is in the Ising model universality class.

After that, we considered the majority-rule model in a fully-connected population with a fraction $d$ of agents with the inflexibility characteristic. In this case, these agents present frozen states and cannot be persuaded to change opinion. In the language of magnetic systems, those special agents behave as the introduction of quenched disorder in the system. We showed that there is a critical fraction $d_{c}=1/2$ above which there is no order in the system, i.e., there is no decision or majority opinion. This order-disorder phase transition is also in the universality class of the mean-field Ising model. We observed consensus in the population only in the absence of inflexible agents, i.e., for $d=0$. In other words, the presence of intrasigents in the population makes the model more realistic, since there is a clear decision in the public debate for $0<d<1/2$. Again, as a test for the universality of the model, we simulated it on triangular and square lattices. In this case, we did not observe a phase transition for both lattices, suggesting that the model with quenched disorder does not undergo a phase transition in small dimensions like $D=2$.

Finally, we considered both effects, independence and inflexibility, in the majority-rule model. In this case we also observed a phase transition at mean-field level, and the critical points $q_{c}$ depend on $f$ and $d$. Consensus states were only obtained for $q=d=0$, and the critical exponents are the same as observed before, i.e., we found again the universality class of the mean-field Ising model. From the phase diagram of this formulation of the model we observed that the increase of the flexibility parameter $f$, related to the independent behavior, makes the ordered phase smaller. 

It was recently discussed that the majority-rule model with limited persuasion can lead to the victory of the initial minority, provided it is sufficiently small \cite{nuno_pmco_minority}. Thus, as a future extension of the present work, it may be interesting to analyze how different initial concentrations of the opinions affect the dynamics, as well as mechanisms of limited persuasion in the majority-rule dynamics.

% ############################################################################

\appendix*
\section{}

\subsection{Analytical calculations: model with independence}

Let us consider the model with independent behavior in the mean-field formulation. Following the approach of Ref. \cite{nuno_indep,biswas}, we computed the stationary order parameter, as well as the critical values $q_{c}(f)$. Let us first define $f_{1}$ and $f_{-1}$ as the stationary probabilities of each possible state ($+1$ or $-1$, respectively). We have to calculate the probability that a given agent suffers the change $+1\to -1$ or $-1\to +1$. We are considering groups of 3 agents, so one can have distinct variations of the magnetization, depending on the states of the 3 agents. For example, the probability to choose at random 3 agents with opinions $o=+1$, i.e, a configuration $(+,+,+)$, is $f_{1}^{3}$. With probability $1-q$ the configuration remains $(+,+,+)$, which does not affect the magnetization of the system. With probability $q\,(1-f)$ the configuration also remains $(+,+,+)$, and with probability $q\,f$ the configuration changes to $(-,-,-)$, which cause a variation of $-6$ in the magnetization. In other words, the magnetization decreases $6$ units with probability $q\,f\,f_{1}^{3}$. One can denote this probability as $r(-6)$, i.e., the probability that the magnetization variation is equal to $-6$. Generalizing, one can define $r(k)$, with $-6\le k \le +6$ in this case, as the probability that the magnetization variation is $k$ after the application of the models' rules. As the order parameter (magnetization) stabilizes in the steady states, we have that its average variation must vanish in those steady states, namely,
\begin{equation} \label{nullshift1}
6\,[r(+6)-r(-6)] + 2\,[r(+2)-r(-2)]=0 \,.
\end{equation}
In this case, we have
\begin{eqnarray} \nonumber
r(+6) &=& f\,q\,f_{-1}^{3} \\ \nonumber
r(-6) &=& f\,q\,f_{1}^{3} \\ \nonumber
r(+2) &=& 3\,(1-q)\,f_{1}^{2}\,f_{-1} + 3\,f\,q\,f_{1}\,f_{-1}^{2} \\ \nonumber
r(-2) &=& 3\,(1-q)\,f_{1}\,f_{-1}^{2} + 3\,f\,q\,f_{1}^{2}\,f_{-1} ~.
\end{eqnarray}
Thus, the null average variation condition Eq. (\ref{nullshift1}) give us
\begin{equation}
(f_{1}-f_{-1})\,[f\,q(f_{1}^{2}+f_{1}\,f_{-1}+f_{-1}^{2}) - (1-q)f_{1}\,f_{-1}+f\,q\,f_{1}\,f_{-1}]=0
\end{equation}
which give us the solution $f_{1}=f_{-1}$ (disordered state) or 
\begin{equation} \label{ap1}
(1-q)\,f_{1}^{2}-(1-q)\,f_{1}+fq=0 ~,
\end{equation}
where we used the normalization condition $f_{1}+f_{-1}=1$. Eq. (\ref{ap1}) give us two solutions for $f_{1}$, and the order parameter can be obtained from $O=|f_{1}-f_{-1}|$, which give us
\begin{equation}
O = \left(1-\frac{4\,q\,f}{1-q}\right)^{1/2} ~.
\end{equation}
\noindent
The critical points $q_{c}$ can be obtained by taking $O=0$, 
\begin{eqnarray} \label{ap2}
q_{c} = \frac{1}{1+4\,f} ~,
\end{eqnarray}
\noindent
that is the Eq. (\ref{eq5}) of the text.

% ##############

\subsection{Analytical calculations: model with inflexibility}

Now we consider the model with inflexibility. Let us now denote the fraction of agents who have opinion $+1$ and are non-inflexibles by $f_{1}$, and similarly for $f_{-1}$. Notice that the total fraction of inflexibles is $d$, and the fraction of inflexibles with opinion $o=+1$ is $d/2$, as well as the fraction of inflexibles with opinion $o=-1$. In this case the normalization condiction becomes \cite{biswas_priv}
\begin{equation} \label{ap3}
f_{1}+f_{-1}=1-d ~,
\end{equation}
\noindent
since the complementary fraction $d$ represents the agents that have frozen states $+1$ or $-1$. Following the same approach of the previous appendix, the null average variation condition becomes 
\begin{equation} \label{nullshift2}
2\,[r(+2)-r(-2)]=0 \,,
\end{equation}
\noindent
where the probabilities $r(k)$ are given by 
\begin{eqnarray} \nonumber
r(+2) &=& 3\,(f_{1}+d/2)^{2}\,f_{-1} \\ \nonumber
r(-2) &=& 3\,(f_{-1}+d/2)^{2}\,f_{1} ~.
\end{eqnarray}
Thus, the null average variation condition Eq. (\ref{nullshift2}) give us
\begin{equation} \label{ap4}
(f_{1}-f_{-1})\,(f_{1}\,f_{-1}-d^{2}/4)=0 ~,
\end{equation}
which give us the solution $f_{1}=f_{-1}$ (disordered state) or 
\begin{equation} \label{ap5}
f_{1}^{2}+(d-1)\,f_{1}+d^{2}/4=0  ~,
\end{equation}
where we used the normalization condition, Eq. (\ref{ap3}). Eq. (\ref{ap5}) give us two solutions for $f_{1}$, and the order parameter can be obtained from $O=|f_{1}-f_{-1}|$, which give us
\begin{equation}
O = (1-2\,d)^{1/2} ~.
\end{equation}
\noindent
The critical point $d_{c}$ can be obtained by taking $O=0$,
\begin{eqnarray} \label{ap6}
d_{c} = \frac{1}{2} ~,
\end{eqnarray}
\noindent
that is the Eq. (\ref{eq8}) of the text.

% ##############

\subsection{Analytical calculations: model with independence and inflexibility}

Now we consider the model with inflexibility and independence. As in the previous section, let us now denote the fraction of agents who have opinion $+1$ and are non-inflexibles by $f_{1}$, and similarly for $f_{-1}$. Notice that the total fraction of inflexibles is $d$, and the fraction of inflexibles with opinion $o=+1$ is $d/2$, as well as the fraction of inflexibles with opinion $o=-1$. The normalization condition is given by Eq. (\ref{ap3})

Following the same approach of the previous appendices, the null average variation condition becomes 
\begin{equation} \label{nullshift3}
6\,[r(+6)-r(-6)]+4\,[r(+4)-r(-4)]+2\,[r(+2)-r(-2)]=0 \,,
\end{equation}
\noindent
where the probabilities $r(k)$ are given by 
\begin{eqnarray} \nonumber
r(+6) &=& f\,q\,f_{-1}^{3}               \\ \nonumber
r(-6) &=& f\,q\,f_{1}^{3}              \\ \nonumber
r(+4) &=& 3\,f\,q\,(d/2)\,f_{-1}^{2}+3\,f\,q\,(d/2)\,f_{-1}^{2}    \\ \nonumber
r(-4) &=&  3\,f\,q\,(d/2)\,f_{1}^{2}+3\,f\,q\,(d/2)\,f_{1}^{2}     \\ \nonumber
r(+2) &=& 3\,f\,q\,(d/2)^{2}\,f_{-1}+3\,(1-q)\,(f_{1}+d/2)^{2}\,f_{-1}+9\,f\,q\,(d/2)^{2}\,f_{-1}+3\,f\,q\,f_{-1}^{2}\,f_{1}     \\ \nonumber
r(-2) &=& 3\,f\,q\,(d/2)^{2}\,f_{1}+3\,(1-q)\,(f_{-1}+d/2)^{2}\,f_{1}+9\,f\,q\,(d/2)^{2}\,f_{1}+3\,f\,q\,f_{1}^{2}\,f_{-1}    \\ \nonumber
\end{eqnarray}
Thus, the null average variation condition Eq. (\ref{nullshift3}) give us
\begin{eqnarray} \nonumber
&&(f_{1}-f_{-1})\,[f\,q\,(f_{1}^{2}+f_{1}\,f_{-1}+f_{-1}^{2})+2\,f\,q\,d(f_{1}+f_{-1})+f\,q\,d^{2}+f\,q\,f_{1}\,f_{-1}  \\
& &\mbox{}-(1-q)(f_{1}\,f_{-1}-d^{2}/4)]=0 ~,
\end{eqnarray}
which give us the solution $f_{1}=f_{-1}$ (disordered state) or 
\begin{equation} \label{ap8}
(1-q)\,f_{1}^{2}+(1-q)\,(d-1)\,f_{1}+[f\,q+(1-q)\,(d^{2}/4)]=0 ~,
\end{equation}
where we used the normalization condition, Eq. (\ref{ap3}). Eq. (\ref{ap8}) give us two solutions for $f_{1}$, and the order parameter can be obtained from $O=|f_{1}-f_{-1}|$, which give us
\begin{equation}
O = [(1-d)^{2}\,(1-\theta)]^{1/2} ~,
\end{equation}
\noindent
where $\theta=\theta(q,d,f)$ is given by
\begin{equation}
\theta=\frac{4}{(1-q)\,(1-d)^{2}}\left\{\,f\,q+(1-q)\,\frac{d^{2}}{4}\right\} ~.
\end{equation}
The critical points $q_{c}$ can be obtained by taking $O=0$,
\begin{eqnarray} \label{ap9}
q_{c} =\frac{1-2\,d}{1+4\,f-2\,d} ~,
\end{eqnarray}
\noindent
that is the Eq. (\ref{eq11}) of the text. Notice that we recover the results (\ref{ap2}) and (\ref{ap6}) for $d=0$ (no inflexibility) and $q_{c}=0$ (no independence), respectively.

% ############################################################################

\section*{Acknowledgments}

The authors acknowledge financial support from the Brazilian scientific funding agencies CNPq and CAPES.

% ############################################################################

\end{document}